\documentstyle[prd,preprint,aps]{revtex}
\begin{document}
\draft
\title{Neutron Beta Decay in a Left-Right Symmetric Model}
\author{A. Guti\'errez-Rodr\'{\i}guez $^{1}$, M. A. Hern\'andez-Ru\'{\i}z $^{2}$, C. L\'opez-Gonz\'alez $^{1}$ and
        M. Maya $^{3}$}
\address{(1) Facultad de F\'{\i}sica, Universidad Aut\'onoma de Zacatecas\\
             Apartado Postal C-580, 98060 Zacatecas, Zacatecas M\'exico.}
\address{(2) Facultad de Ciencias Qu\'{\i}micas, Universidad Aut\'onoma de Zacatecas\\
             C\'odigo Postal 98068, Zacatecas, Zacatecas M\'exico.}
\address{(3) Facultad de Ciencias F\'{\i}sico Matem\'aticas, Universidad Aut\'onoma de Puebla\\
             Apartado Postal 1364, 72000, Puebla, Puebla M\'exico.}
\date{\today}
\maketitle

\begin{abstract}

We start from a Left-Right Symmetric Model and we analyze the beta decay of
the free neutron $n\rightarrow p+e^{-}+\bar{\nu}_{e}$. We applied this model
to incorporate the right currents, whereby we propose an amplitude whose
leptonic part contains the parameter $\lambda $ defined as left-right
asymmetry parameter which measures the parity violation. The analysis
consists of seeing how the spectrum of energy of the electrons, the total
rate of decay, and the lifetime $\tau $ of the neutron are affected by the
left-right asymmetry parameter, besides taking into account corrections of
mass, that is, $m_{e}\neq 0$, $m_{p}\neq m_{n}$, and the recoil of the proton.
\end{abstract}



\narrowtext

\section{Introduction}

Matter is composed of atoms which, although they are of very different
classes, contain the same basic constituents: the proton, the neutron and
the electron. Of these three, two (the proton and the neutron) are only in
the nucleus and because of this they are usually called nucleones.

The neutron is a particle with charge zero, mass in rest of $939.565330\pm
0.000038$ $MeV$ \cite{Data}, and spin $\hbar /2$. The neutron is a fermion,
which decays or disintegrates for weak interaction in: $n\rightarrow p+e^{-}+\bar{\nu}_{e}$
\cite{Data,Chadwick}, that is to say, the neutron becomes a
proton emitting an electron plus an antineutrino of the electron. The
experimental confirmation of the discovery of the neutron was made in 1932
by James Chadwick \cite{Chadwick}.

The free neutron is unstable and this instability results in that the
neutron decays. However, in the nuclei it can remain without disintegrating.
This is due to the fact that inside the nucleus, the protons and neutrons
are connected by the nuclear forces, that is to say, by means of a strong
interaction.

The process of beta decay of the neutron is: $n\rightarrow p+e^{-}+\bar{\nu}_{e}$.
The theory of Fermi of beta decay is used currently, although with
some changes in weak processes to low energy, now called the theory V-A. For
weak processes to high energy, the correct theory is the electroweak theory
of Weinberg, Salam and Glashow \cite{Weinberg}, of which the theory V-A is a
limit.

One of the important characteristics of the decay of the neutron is that
this process has current interest in nuclear physics \cite{Amaldi} and in
cosmological theories \cite{Amaldi,Byrne,Dubbers} since it is necessary to
have complete theoretical results of the total rate of decay and thus, more
and more precise experimental results of the lifetime $\tau $ of the neutron 
\cite{Data,Byrne,Dubbers,Yerozolimsky}.

The total rate of decay of the neutron is an important quantity for its
cosmological impact on the synthesis of light elements \cite{Byrne,Dubbers}
and because, combined with the asymmetry of the electron, it provides in the
standard model a direct determination of the coupling constants vectorial
semileptonic $(F_{V})$ and axial-vector $(G_{A})$ \cite{Data,Byrne,Dubbers,Yerozolimsky}.

The mensurations of the beta decay of the neutron provide important
information about the strong and electroweak interactions. The decay of the
free neutron is distinguished among the beta decay, because (i) it is
theoretically quite pure (the theoretical uncertainties are small); (ii) the
free neutrons can produce abundantly without expensive accelerators, and the
detectors of the products of decay of the neutron are also relatively
economic.

The experiments of the decay of the free neutron make possible the precise
and exact determination of the following important parameters of the
standard model of the elementary particles: the constants of coupling weak
vector $(F_{V})$ and axial-vector $(G_{A})$, and the elements $V_{ud}$ of
the matrix of Kobayashi-Maskawa (KM). The value of $F_{V}$ is important for
the test of the conservation of the hypothesis of the vectorial current;
$V_{ud}$ is necessary to prove the unitary nature of the matrix KM; the
reason $\alpha =G_{A}/F_{V}$ helps to understand the fundamental interaction
of the first quarks generation. $\alpha$ can be determined for diverse
forms and independently of mensurations of the decay of the neutron, and the
comparison of the several results of $\alpha$ provides an important test of
the phenomenological model V-A (the approach to low energy of the part of
the standard model of the weak interaction of the charged current). The
precise value of $\alpha$ is important for the calculation of cross
sections of the weak hadronic interaction, for applications in astrophysics,
cosmology big bang, solar physics and neutrinos detection (see the articles 
\cite{Byrne,Dubbers,Mampe,Gluck,Deutsch} and references therein).

There are many extensions of the standard model that predict measured
effects of deviations of the standard model in decays of the neutron. One of
the most popular is the model with left-right symmetry, with the norm group
$SU(2)_{L}\times SU(2)_{R}\times U(1)$ and currents charged with right
helicity \cite{Senjanovic,Carnoy}.

The purpose of this work is to carry out an analysis of the decay of the
free neutron, in the context from a model with left-right symmetry
\cite{Huerta}, for this we start from an extension of
the electroweak model applied to the baryons decay \cite{Augusto}. This
model contains the parameter $\lambda $ defined as parameter of left-right
asymmetry, which measures the parity violation. We apply this theory to
incorporate the right currents, for which we propose an amplitude whose
leptonic part is $V+\lambda A$, with $\lambda =-1$ for left currents and
$\lambda =1$ for right currents. We also consider the mass of the electron
different from zero $(m_{e}\neq 0)$, the recoil of the proton and the
difference of masses among the proton and the neutron $(m_{p}\neq m_{n})$.

The central point of ours result is that we have considered an interaction
point of four fermions, and we have considered the proton and the neutron as
if they were simply point particles. The most realistic thing is to consider
that the proton and the neutron possess structure which is coupled to the
weak boson $W$, and to express the transition amplitude in terms of several
form factors whose structure is limited by the covariance of Lorentz. However,
this case goes beyond the purpose of this work and thus is not considered
here.

This work is structured in the following way: In Section II, we consider the
model with left-right symmetry, which is the point of departure of this
work. Section III contains our analytic calculations of the decay of the
neutron, and finally, Section IV contains our results and conclusions. An
appendix is also included, which contains the calculations of the total rate
of decay for the case when the mass of the electron is different from zero,
that is, $y=\frac{m_{e}}{m_{n}}\neq 0$.

\section{Theoretical Model}

In the theory of Fermi V-A the transition amplitude for the beta decay of
the neutron is given by the expression

\begin{equation}
{\cal M}=\frac{G_F}{\sqrt{2}}[\bar u_p \gamma^\alpha(F_V+G_A\gamma_5)u_n]
[\bar u_e \gamma_\alpha(1-\gamma_5)v_\nu],
\end{equation}

\noindent where $F_V$ and $G_A$ are produced by the strong interaction of
the proton and of the neutron.

Experimentally it is found that the coefficient $F_{V}$ of the vector it is
$F_{V}=1$ and the coefficient $G_{A}$ of the axial-vector is $G_{A}=1.2670\pm 0.0035$ \cite{Data}.

Although the experiments carried out up to now indicate in a definitive
manner that the emitted neutrinos are of mainly negative helicity (which is
known as left currents), there is not a fundamental theoretical reason for
the same happening to higher energy.

The electroweak model takes as a fact that there are only neutrinos of
negative helicity and, as it was commented previously, their limit to low
energy is the theory V-A. Other physicists extend the electroweak theory
building a model with left-right symmetry \cite{Beg}, that is, incorporating
neutrinos of positive helicity (which is known as right currents).

The focus that we use here to include right currents is exposed by R. Huerta
in the Ref. \cite{Huerta}, and later is used by A. Garc\'{\i}a {\it et al.} 
\cite{Augusto}, in semileptonics decay of baryons.

We take as starting point the amplitude

\begin{equation}
{\cal M}=\frac{G}{\sqrt{2}}
[aJ^l_LJ^h_L+b(J^l_LJ^h_R+J^l_RJ^h_L)+cJ^l_RJ^h_R],
\end{equation}

\noindent where $a$, $b$, and $c$ contain the parameters of the electroweak
model with left-right symmetry, which is not dealt with here, but we can say
that it is the version to low energy of the electroweak symmetrical model.

The amplitude (2) is for the decay $A\rightarrow B+e^{-}+\bar{\nu}_{e}$
where $A$ and $B$ are baryons. Using the notation: $\mid 1>$ for the
neutrino, $\mid 2>$ for the electron and $\mid A>$, $\mid B>$ for the
baryons, it has that the left leptonic part is:

\[
J^l_L=<2\mid V-A\mid 1>, 
\]

\noindent which contains the neutrinos of negative helicity; the right
leptonic part is

\[
J^l_R=<2\mid V+A\mid 1>, 
\]

\noindent while the left baryonic part is:

\[
J^h_L=<B\mid F_L V+G_L A\mid A>, 
\]

\noindent and the right baryonic part

\[
J^h_R=<B\mid F_R V+G_R A\mid A>, 
\]

\noindent where $F_L$, $F_R$, $G_L$, $G_R$ are induced form factors for the
strong interaction. Substituting the expressions of the currents in (2)

\begin{eqnarray}
{\cal M}&=&\frac{G_F}{\sqrt{2}} [a <2\mid V-A\mid 1><B\mid F_LV+G_LA\mid A> \nonumber \\
&& +b <2\mid V-A \mid 1><B\mid F_R V+G_RA\mid A>  \nonumber \\
&& +b <2\mid V+A\mid 1><B\mid F_LV+G_LA\mid A> \\
&& +c <2\mid V+A\mid 1><B\mid F_RV+G_RA\mid A>],  \nonumber
\end{eqnarray}

\noindent and after regrouping appropriately and defining

\begin{eqnarray}
{\cal P}&=&aF_L+bF_R+bF_L+cF_R,  \nonumber \\
{\cal Q}&=&-aF_L-bF_R+bF_L+cF_R,  \nonumber \\
{\cal R}&=&aG_L+bG_R+bG_L+cG_R, \\
{\cal S}&=&-aG_L-bG_R+bG_L+cG_R,  \nonumber
\end{eqnarray}

\noindent we obtain

\begin{equation}
{\cal M}=\frac{G_F}{\sqrt{2}} [<2\mid V+\frac{\cal Q}{\cal P}A\mid 1><B\mid {\cal P}V\mid A>+
<2\mid V+\frac{\cal S}{\cal R}A\mid 1><B\mid {\cal R}A\mid A>],
\end{equation}

\noindent where

\begin{eqnarray}
\lambda=\frac{\cal Q}{\cal P}=\frac{-aF_L-bF_R+bF_L+cF_R}{aF_L+bF_R+bF_L+cF_R}, \nonumber \\
\lambda ^{\prime}= \frac{\cal S}{\cal R}=\frac{-aG_L-bG_R+bG_L+cG_R}{aG_L+bG_R+bG_L+cG_R}.
\end{eqnarray}

As the strong interaction is invariant under parity, we can suppose that

\[
F_L=F_R, \hspace*{5mm} G_L=G_R \hspace*{3mm}{\mbox {and then}} \hspace{5mm}
\lambda=\lambda^{\prime}= \frac{-a+c}{a+2b+c}, 
\]

\noindent and furthermore define

\begin{eqnarray*}
F\equiv {\cal P}&=&aF_L+bF_R+bF_L+cF_R=(a+2b+c)F_L, \\
G\equiv {\cal R}&=&aG_L+bG_R+bG_L+cG_R=(a+2b+c)G_L. \\
\end{eqnarray*}

Substituting these expressions and ordering appropriately, we obtain the
amplitude of decay for the model with left-right symmetry

\begin{equation}
{\cal M}=\frac{G_F}{\sqrt{2}} [<2\mid V+\lambda A\mid 1><B\mid FV+GA\mid A>].
\end{equation}

\noindent In this amplitude, the effects of the currents are already
included in $\lambda $ and in the form factors $F$ and $G$.

\section{Decay of the Neutron}

As we already mentioned in the introduction, the neutron is a particle with
charge zero, spin $\hbar /2$ and mass in rest of $939.565330\pm 0.000038$ $MeV$.
This particle decays or disintegrates for weak interaction in:

\begin{equation}
n\rightarrow p+e^-+\bar \nu_e,
\end{equation}

\noindent as indicated in the diagram of Feynman, Fig. 1.

\subsection{Amplitude of Transition}

The amplitude of decay of the neutron is

\begin{equation}
{\cal M}=\frac{G_F}{\sqrt{2}} [\bar u_p\gamma^\alpha(1-\rho\gamma_5)u_n]
[\bar u_e\gamma_\alpha(1+\lambda\gamma_5)v_\nu].
\end{equation}

The coefficient $\lambda $ that appears in the leptonic part of the Eq. (9),
is what we call the coefficient of left-right asymmetry, which contains
information on the effects of the theory V+A (right currents). This
coefficient does not appear in the common literature and it is introduced to
incorporate the theory V-A and part of V+A, and, as already commented, is
the starting point of this work.

To determine the square of the amplitude of transition of our process, we
should determine the conjugated complex firstly of (9)

\begin{equation}
{\cal M^*} = \frac{G_F}{\sqrt{2}} [\bar v_\nu\gamma_\beta(1+\lambda\gamma_5)u_e]
[\bar u_n\gamma^\beta(1-\rho\gamma_5)u_p].
\end{equation}

Now, of the Eqs. (9) and (10), and after applying some of the theorems of
traces of the Dirac matrix , we have that the square of the amplitude of
transition is

\begin{eqnarray}
\sum_s\mid {\cal M}\mid^2&=&\frac{G^2_F}{2}[32(1+\rho^2)(1+\lambda^2)\{(p_p
\cdot p_e)(p_n\cdot p_\nu) +(p_p\cdot p_\nu)(p_n\cdot p_e) \}  \nonumber \\
&& +32m_\nu m_e(1+\rho^2)(1-\lambda^2)(p_n\cdot p_e)
+128\lambda\rho\{(p_p\cdot p_\nu)(p_n\cdot p_e)-(p_p\cdot p_e)(p_n\cdot
p_\nu) \}  \nonumber \\
&& -32m_nm_p(1-\rho^2)(1+\lambda^2)(p_\nu \cdot p_e)-64m_\nu
m_em_pm_n(1-\rho^2)(1-\lambda^2)].
\end{eqnarray}

To simplify the calculations, we move to the system center of masses of the
neutron where ${\bf p}_{n}=0$, so that the four-momentum products are:

\begin{eqnarray}
p_p\cdot p_e&=&E_pE_e-{\bf p}_p\cdot {\bf p}_e,  \nonumber \\
p_n\cdot p_\nu&=&m_nE_\nu,  \nonumber \\
p_p\cdot p_\nu&=&E_pE_\nu-{\bf p}_p\cdot {\bf p}_\nu,  \nonumber \\
p_n\cdot p_e&=&m_nE_e, \\
p_n\cdot p_p&=&m_nE_p,  \nonumber \\
p_\nu\cdot p_e&=&E_\nu E_e-{\bf p}_\nu\cdot {\bf p}_e.  \nonumber
\end{eqnarray}

Up until now, there are no precise results on the mass of the neutrino of
the electron, it is only known that it is very small compared with the mass
of the electron $(m_{\nu }<<m_{e})$ so that we can suppose that $m_{\nu }=0$,
and of the energy-momentum relativistic relationship for the neutrino we
have that $E_{\nu }^{2}=m_{\nu }^{2}+p_{\nu }^{2}$ where $E_{\nu }=p_{\nu }$,
which simplifies the calculations.

\noindent To simplify even more the expression (11), we define the following
which will be later of great utility

\[
\beta_e=\frac{\mid {\bf p}_e\mid}{E_e}, \hspace*{5mm} ({\bf \hat p}_e\cdot 
{\bf \hat p}_\nu)=x_{e\nu}=x, \hspace*{5mm} A_\rho=(1+\rho^2), 
\]

\[
A_\lambda=(1+\lambda^2), \hspace*{5mm} B_\rho=-2\rho, \hspace*{5mm}
B_\lambda=2\lambda, \hspace*{5mm} C_\rho=(1-\rho^2), 
\]

\noindent thus, we have finally that the square of the transition amplitude
for our process is

\begin{equation}
\sum_s\mid {\cal M} \mid ^2=16G^2_F
m_n[M_1E_\nu+M_2E^2_\nu+(N_1E_\nu+N_2E^2_\nu)({\bf \hat p}_e\cdot {\bf \hat p}_\nu)].
\end{equation}

\noindent This expression contains the details of the interaction, that is,
it describes the dynamics of the process.

\noindent In (13) we have defined

\begin{eqnarray}
M_1&=&A_\rho A_\lambda(2E_e m_n-E^2_e-m^2_e)-A_\lambda C_\rho E_e m_p+B_\rho
B_\lambda(E^2_e-m^2_e),  \nonumber \\
M_2&=&(-A_\rho A_\lambda-B_\rho B_\lambda) E_e,  \nonumber \\
N_1&=&(A_\rho A\lambda E_e+A_\lambda C_\rho m_p-B_\rho B_\lambda
E_e)E_e\beta_e, \\
N_2&=&(A_\rho A_\lambda+B_\rho B_\lambda)E_e\beta_e.  \nonumber
\end{eqnarray}

\subsection{Differential Decay Rate}

Our following step, now that we know the square of the transition amplitude
Eq. (13), is to calculate the differential rate of decay of the neutron.

The expression for the differential rate of decay for a particle that decays
in three is expressed by \cite{Data}:

\begin{equation}
d\Gamma=(2\pi)^4\mid <abc\mid \tau\mid A> \mid^2\frac{1}{2E_A}
\delta^4(p_A-p_a-p_b-p_c)\frac{d^3{\bf p}_a d^3{\bf p}_b d^3{\bf p}_c}
{(2\pi)^3 2E_a(2\pi)^3 2E_b(2\pi)^3 2E_c},
\end{equation}

\noindent where $\mid A>$ is the initial state of the system, $<abc\mid $ is
the final state and $\tau $ is the operator that makes the transition. For
our case $<pe^{-}\bar{\nu}\mid \tau \mid n>={\cal M}$.

Applying the expression (15) to our process we have that

\begin{equation}
d\Gamma=\frac{\sum_s\mid {\cal M}\mid^2}{16(2\pi)^5m_n}\frac{d^3{\bf p}_e d^3
{\bf p}_\nu d^3{\bf p}_p}{E_e E_\nu E_p}\delta(E_n-E_e-E_\nu-E_p)\delta^3
({\bf p}_n - {\bf p}_e - {\bf p}_\nu - {\bf p}_p),
\end{equation}

\noindent in this expression, the square of the transition amplitude
$\sum_{s}\mid {\cal M}\mid ^{2}$ provides information on the details of the
interaction, the Dirac delta function implies conservation of energy-moment,
and the other terms describe the kinematics of the process, which are
standard for all the reactions.

The order in which (16) should be integrated will depend on the physical
quantities wanted, as well as on the ease of carring out the integrations.

The first aspect that we want is to calculate the spectrum of energy of the
electron, reason for which we will integrate firstly with regard to the
moment of the proton ${\bf p}_{p}$:

\begin{equation}
d\Gamma=\frac{\sum_s\mid {\cal M}\mid^2}{16(2\pi)^5m_n} \frac{d^3{\bf p}_e
d^3{\bf p}_\nu}{E_e E_\nu E_p}\delta(E_n-E_e-E_\nu-E_p).
\end{equation}

Now we integrate with regard to the moment of the neutrino; for this we use
spherical coordinates, and from the energy-momentum relativist relationship,
we have that

\[
d^3{\bf p}_\nu=p_\nu E_\nu d\Omega_\nu dE_\nu, 
\]

\noindent then

\begin{equation}
d\Gamma=\frac{\sum_s\mid {\cal M}\mid^2}{16(2\pi)^5m_n} \frac{d^3{\bf p}_e
p_\nu}{E_eE_p}d\Omega_\nu\delta(E_n-E_e-E_\nu-E_p)dE_\nu.
\end{equation}

To integrate the Eq. (18) with regard to the energy of the neutrino $E_\nu$,
we use the following property of the delta of Dirac:

\[
\int G(E_\nu)\delta[f(E_\nu)] dE_\nu=\frac{G(E^0_\nu)} {\mid
f^{\prime}(E^0_\nu)\mid}, 
\]

\noindent where we have defined

\[
G(E_\nu)=\frac{\sum_s\mid {\cal M}\mid^2}{16(2\pi)^5m_n} \frac{d^3{\bf p}_e
p_\nu}{E_eE_p}d\Omega_\nu, 
\]

\[
f(E_\nu)=E_n-E_e-E_\nu-E_p, \hspace*{5mm} f(E^0_\nu)=E_n-E_e-E_\nu-E_p=0, 
\]

\noindent Here, $E_{\nu }^{0}$ corresponds to the real energy of the
neutrino of the electron, energy for which the function $f(E_{\nu })$ is
made zero. Then we have that the Eq. (18) takes the form

\[
d\Gamma=\frac{\sum_s\mid {\cal M}\mid^2}{16(2\pi)^5m^2_n} \frac{d^3{\bf p}_e}
{E_e}p_\nu d\Omega_\nu[\frac{1}{1-\frac{E_e}{m_n}+\frac{p_e}{m_n}x}],
\]

\noindent where we have defined $({\bf \hat{p}}_{e}\cdot {\bf \hat{p}}_{\nu
})=x$ which is the angle formed between the direction of the electron and
the neutrino.

\noindent Now we define the following again $a=1-\frac{E_{e}}{m_{n}}$, 
\hspace{3mm} $b=\frac{p_{e}}{m_{n}}$, and later we integrate with regard to
the solid angle obtaining

\begin{equation}
d\Gamma=\frac{G^2_F}{(2\pi)^4m_n}\frac{d^3{\bf p}_e}{E_e}\frac{p_\nu}{(a+bx)}
[M_1E_\nu +M_2E^2_\nu +(N_1E_\nu +N_2E^2_\nu)x]dx.
\end{equation}

\noindent To integrate with regard to the variable $x$ we should express
$E_{\nu }=p_{\nu }$ as a function of $x$, which is achieved using
energy-momentum conservation. We have this way that

\begin{equation}
d\Gamma=\frac{G^2_F}{(2\pi)^4m_n}\frac{d^3{\bf p}_e}{E_e}
[M_1\frac{(E_m-E_e)^2}{(a+bx)^3}+M_2\frac{(E_m-E_e)^3}{(a+bx)^4}
+N_1\frac{(E_m-E_e)^2}{(a+bx)^3}x +N_2\frac{(E_m-E_e)^3}{(a+bx)^4}x]dx,
\end{equation}

\noindent where

\[
E_\nu=p_\nu=\frac{(E_m-E_e)} {(a+bx)}, \hspace{1cm} E_m=\frac{m^2_n - m^2_p
+ m^2_e}{2m_n}, 
\]

\noindent and the last equation corresponds to the expression for the
maximum energy of the electron.

\noindent The following is integrated with regard to the variable $x$:

\begin{equation}
d\Gamma=\frac{G^2_F}{(2\pi)^4m_n}\frac{d^3{\bf p}_e}{E_e}
[M_1(E_m-E_e)^2I_1+M_2(E_m-E_e)^3I_2+N_1(E_m-E_e)^2I_3+N_2(E_m-E_e)^3I_4],
\end{equation}

\noindent where

\[
I_1=\int^1_{-1}\frac{dx}{(a+bx)^3}, \hspace*{3mm} I_2=\int^1_{-1}\frac{dx}{(a+bx)^4},
\hspace*{3mm} I_3=\int^1_{-1}\frac{xdx}{(a+bx)^3}, \hspace*{3mm}
I_4=\int^1_{-1}\frac{xdx}{(a+bx)^4}, 
\]

\noindent and we also define

\[
y=\frac{m_e}{m_n}, \hspace*{5mm} z=\frac{m_p}{m_n}, \hspace*{5mm} \epsilon=\frac{E_e}{m_n},
\hspace*{5mm} \epsilon_m=\frac{(1-z^2)} {2}, \hspace*{5mm}
\epsilon^{\prime}_m=\frac{E_m}{m_n}, \hspace*{3mm} 
\]

\[
\eta=\frac{p_e}{m_n}=\sqrt{\epsilon^2-y^2}, \hspace*{5mm} a_0=1+y^2,
\hspace*{5mm} b_0=-2, \hspace*{5mm} c_0=\frac{a_0}{b_0}=\beta.
\]

\noindent After evaluating the integral explicitly and rewriting in terms of
the quantities defined above, as well as the coefficients $M_{1}$, $M_{2}$,
$N_{1}$, $N_{2}$, we obtain

\[
I_1=\frac{2(1-\epsilon)} {(a_0+b_0\epsilon)^2}, \hspace*{3mm} I_2=
\frac{2(3-6\epsilon+4\epsilon^2-y^2)}{3(a_0+b_0\epsilon)^3}, \hspace*{3mm}
I_3=-\frac{2\eta}{(a_0+b_0\epsilon)^2}, \hspace*{3mm} I_4=-\frac{8(1-\epsilon)\eta}{3(a_0+b_0\epsilon)^3}, 
\]

\noindent where now:

\begin{eqnarray}
M_1&=&[A_\rho A_\lambda (2\epsilon-\epsilon^2-y^2)-A_\lambda C_\rho
z\epsilon +B_\rho B_\lambda (\epsilon^2 - y^2)] m^2_n,  \nonumber \\
M_2&=&[-A_\rho A_\lambda \epsilon-B_\rho B_\lambda \epsilon]m_n,  \nonumber
\\
N_1&=&[A_\rho A_\lambda \epsilon+A_\lambda C_\rho z - B_\rho B_\lambda
\epsilon]\eta m^2_n,  \nonumber \\
N_2&=&[A_\rho A_\lambda +B_\rho B_\lambda]\eta m_n.  \nonumber
\end{eqnarray}

\noindent To integrate with regard to the moment of the electron, we express
the differential of volume $d^{3}{\bf p}_{e}$ in spherical coordinates

\[
d^3{\bf p}_e=p^2_e dp_e d\Omega \Rightarrow p^2_edp_e4\pi=m^3_n\eta\epsilon
d\epsilon 4\pi, 
\]

\noindent obtaining the following

\begin{eqnarray}
d\Gamma&=&\frac{4G^2_F m^5_n(\epsilon_m^{\prime}-\epsilon)^2}{(2\pi)^3(a_0+b_0\epsilon)^3}\eta d\epsilon
[(- A_\rho A_\lambda-B_\rho B_\lambda)a_0y^2+A_\lambda C_\rho za_0y^2 +\frac{4}{3}(A_\rho
A_\lambda+B_\rho B_\lambda)\epsilon^{\prime}_my^2 \nonumber \\
&& +\{(-A_\rho A_\lambda-B_\rho B_\lambda)(b_0-a_0)y^2+(2A_\rho
A_\lambda-A_\lambda C_\rho z)a_0 +\frac{1}{3}(-A_\rho A_\lambda-B_\rho
B_\lambda)(3-y^2)\epsilon^{\prime}_m  \nonumber \\
&& +A_\lambda C_\rho zb_0y^2+(A_\rho A_\lambda-B_\rho B_\lambda)a_0y^2 +
\frac{4}{3}(-A_\rho A_\lambda-B_\rho B_\lambda)(1+\epsilon^{\prime}_m)y^2 \}
\epsilon  \nonumber \\
&& +\{(A_\rho A_\lambda+B_\rho B_\lambda)b_0 y^2+(2A_\rho
A_\lambda-A_\lambda C_\rho z)(b_0-a_0) +(-A_\rho A_\lambda+B_\rho
B_\lambda)a_0 \\
&& +2(A_\rho A_\lambda+B_\rho B_\lambda)\epsilon^{\prime}_m +\frac{1}{3}
(A_\rho A_\lambda+B_\rho B_\lambda)(3-y^2) - A_\lambda C_\rho za_0 +(A_\rho
A_\lambda-B_\rho B_\lambda)b_0 y^2  \nonumber \\
&& +\frac{4}{3}(-A_\rho A_\lambda-B_\rho B_\lambda)(\epsilon^{\prime}_m-y^2)
\} \epsilon^2 +\{-(2A_\rho A_\lambda-A_\lambda C_\rho z)b_0+(-A_\rho
A_\lambda+B_\rho B_\lambda)(b_0-a_0)  \nonumber \\
&& +\frac{1}{3}(-A_\rho A_\lambda-B_\rho
B_\lambda)(6+4\epsilon^{\prime}_m)-A_\lambda C_\rho zb_0+(-A_\rho
A_\lambda+B_\rho B_\lambda)a_0  \nonumber \\
&& +\frac{4}{3}(-A_\rho A_\lambda-B_\rho B_\lambda)(-1-\epsilon^{\prime}_m)
\} \epsilon^3].  \nonumber
\end{eqnarray}

As a first approach, we will take as zero the mass of the electron in the
previous expression, that is, $y=m_{e}/m_{n}=0$ since, compared with other
terms, their contribution is very small. With this approach and after making
some pertinent arrangements, we find that the expression for the spectrum of
energy of the electrons is

\begin{equation}
\frac{d\Gamma}{d\epsilon}=\frac{2G^2_F m^5_n}{(2\pi)^3}\frac{\eta}{b^3_0(c_0+\epsilon)^3}
\sum^5_{n=1}F_n\epsilon^n,
\end{equation}

\noindent where:

\begin{eqnarray}
F_1&=&Q\epsilon^2_m,  \nonumber \\
F_2&=&R\epsilon^2_m-2Q\epsilon_m,  \nonumber \\
F_3&=&S\epsilon^2_m-2R\epsilon_m+Q,  \nonumber \\
F_4&=&-2S\epsilon_m+R,  \nonumber \\
F_5&=&S,  \nonumber
\end{eqnarray}

\noindent with

\begin{eqnarray}
Q&=&2A_\rho A_\lambda-A_\lambda C_\rho z-(A_\rho A_\lambda+B_\rho B_\lambda)\epsilon_m,  \nonumber \\
R&=&-6A_\rho A_\lambda+2B_\rho B_\lambda+2A_\lambda C_\rho z+\frac{2}{3}
(A_\rho A_\lambda+B_\rho B_\lambda)\epsilon_m,  \nonumber \\
S&=&\frac{16}{3}A_\rho A_\lambda-\frac{8}{3}B_\rho B_\lambda.  \nonumber
\end{eqnarray}

\subsection{Total Decay Rate}

To determine the total rate of decay of the neutron we integrate the
expression (23) with regard to the variable $\epsilon$, that is,

\begin{equation}
\Gamma=\frac{2G^2_F m^5_n}{(2\pi)^3b^3_0}\sum^5_{n=1}F_nI_n,
\end{equation}

\noindent where

\[
I_n=\int^{\epsilon_1}_{\epsilon_0}\eta\frac{\epsilon^n}{(c_0+\epsilon)^3}
d\epsilon, \hspace*{8mm} n=1,2,3,.. ,5, \hspace*{5mm} \epsilon_0=0, \hspace*{5mm}
\epsilon_1=\epsilon_m,
\]

\noindent This type of integral is not difficult to solve, since with an
appropriate change of variables, these are solved quickly.

Explicitly the sum of the Eq. (24) is

\begin{eqnarray}
\sum^5_{n=1}F_nI_n&=&(\frac{1}{4}F_1+\frac{1}{8}F_2+\frac{1}{16}F_3
+\frac{1}{32}F_4+\frac{1}{64}F_5)J_{-3} +(F_1+\frac{3}{4}F_2+\frac{1}{2}F_3+\frac{5}{16}F_4
+\frac{3}{16}F_5)J_{-2}  \nonumber \\
&& +(F_1+\frac{3}{2}F_2+\frac{3}{2}F_3+\frac{5}{4}F_4+\frac{15}{16}
F_5)J_{-1} +(F_2+2F_3+\frac{5}{2}F_4+\frac{5}{2}F_5)J_0 \\
&& +(F_3+\frac{5}{2}F_4+\frac{15}{4}F_5)J_1 +(F_4+3F_5)J_2
+F_5J_3=\sum^3_{n=-3}G_nJ_n,  \nonumber
\end{eqnarray}

\noindent where the $J_{n}$ are the aforementioned integrals, which are

\[
J_{-3}=\int^{\alpha_1}_{\alpha_0}\frac{1}{\alpha^3}d\alpha=\frac{2\epsilon^2_m-2\epsilon_m}{(\epsilon_m-1/2)^2},
\hspace*{1cm} J_1=\int^{\alpha_1}_{\alpha_0}\alpha d\alpha=\frac{1}{2}(\epsilon^2_m-\epsilon_m), 
\]
\[
J_{-2}=\int^{\alpha_1}_{\alpha_0}\frac{1}{\alpha^2}d\alpha=-\frac{2\epsilon_m
}{(\epsilon_m-1/2)}, \hspace*{1cm} J_2=\int^{\alpha_1}_{\alpha_0}\alpha^2
d\alpha=\frac{1}{3}\epsilon^3_m-\frac{1}{2}\epsilon^2_m+\frac{1}{4}
\epsilon_m, 
\]
\[
J_{-1}=\int^{\alpha_1}_{\alpha_0}\frac{1}{\alpha}d\alpha=ln(1-2\epsilon_m), 
\hspace*{1cm} J_3=\int^{\alpha_1}_{\alpha_0}\alpha^3 d\alpha=\frac{1}{4}
\epsilon^4_m-\frac{1}{3}\epsilon^3_m+\frac{3}{8}\epsilon^2_m-\frac{1}{8}
\epsilon_m, 
\]
\[
J_{0}=\int^{\alpha_1}_{\alpha_0}d\alpha=\epsilon_m \newline
, 
\]

\noindent and their respective coefficients $G_n$:

\begin{eqnarray}
G_{-3}&=&-\frac{1}{6}(A_\rho A_\lambda+B_\rho B_\lambda)(\epsilon_m - \frac{1}{2})^3,  \nonumber \\
G_{-2}&=&-\frac{1}{2}(A_\rho A_\lambda +B_\rho B_\lambda)(\epsilon_m - \frac{1}{2})^2+A_\lambda C_\rho z\frac{1}{2}(\epsilon_m-\frac{1}{2})^2,  \nonumber\\
G_{-1}&=&A_\rho A_\lambda(2\epsilon^2_m-2\epsilon_m+\frac{1}{2}) +B_\rho B_\lambda(0)+A_\lambda C_\rho z(2\epsilon^2_m-3\epsilon_m+1),  \nonumber \\
G_0&=&A_\rho A_\lambda(\frac{2}{3}\epsilon^3_m +4\epsilon_m^2 -7\epsilon_m +\frac{7}{3}) +B_\rho B_\lambda (\frac{2}{3}\epsilon^3_m -4\epsilon^2_m
+5\epsilon_m-\frac{5}{3})  
+A_\lambda C_\rho z(2\epsilon^2_m-6\epsilon_m+3),  \nonumber \\
G_1&=&A_\rho A_\lambda(4\epsilon^2_m -14\epsilon_m+7)+B_\rho
B_\lambda(-4\epsilon^2_m +10\epsilon_m -5)+A_\lambda C_\rho
z(-4\epsilon_m+4), \\
G_2&=&A_\rho A_\lambda(-10\epsilon_m+10)+B_\rho B_\lambda(6\epsilon_m
-6)+2A_\lambda C_\rho z,  \nonumber \\
G_3&=&\frac{16}{3}A_\rho A_\lambda - \frac{8}{3}B_\rho B_\lambda.  \nonumber
\end{eqnarray}

As we already have explicitly the integrals $J_{n}$ with their respective
coefficients $G_{n}$, the following is to substitute both results in the Eq.
(24), and after a simple algebra we obtain the total rate of decay of the
neutron:

\begin{equation}
\Gamma=\frac{G^2_Fm^5_n}{60\pi^3}\epsilon^5_m(1+\lambda^2)[(1+\rho^2)(1+\epsilon_m+....)
+z(1-\rho^2)(-\frac{1}{2}-\epsilon_m +....)].
\end{equation}

This result is consistent with those of the common literature, those that
only contain left currents. The new contribution in the Eq. (27) is the
parameter $\lambda $ that, as we have already mentioned, gives us
information on those effects of the right currents $(V+A)$, and $\epsilon
_{m}$ (where $\epsilon _{m}^{2}$ and $\epsilon _{m}^{3}$ have been rejected
because their contribution is very small) which is the most important term
of correction. The numeric value of this term is presented in the part
corresponding to the results and conclusions.

An immediate consequence of the Eq. (27) is that we can calculate
immediately the lifetime $\tau $ of the neutron, that is, the required time
for a certain sample of particles or nuclei to disintegrate until decreasing
by half.

\section{Results and Conclusions}

In this part we present our results and conclusions of the decay of the free
neutron. Firstly, we have the corresponding result for the spectrum of
energy of the electrons that, as is known, during the beta disintegration,
the energy spectra of the electrons are always continuous.

The physical meaning of the Eq. (23), that corresponds to the spectrum of
energy of the electrons, is that this expression determines the total number
of electrons that leave with a certain energy (or the total probability of
emission of an electron).

An important point here is to see the influence of the parameter of
left-right asymmetry $\lambda $ in the spectrum or distribution of energy of
the electrons, thus, a group of figures for different values of this
parameter is presented.

Fig. 1 corresponds to $\lambda =-1$, that is to say, for left currents, Fig.
2, for $\lambda =1$ right currents, while in Fig. 3, several curves for
different values of the parameter of asymmetry $\lambda $ are superimposed.

Comparing these figures, it is clear that the parameter of asymmetry
$\lambda$ does not modify the curve of energy distribution, and only drops
or goes up depending on the value of $\lambda $, as is observed more clearly
in Fig. 3; but we consider that this parameter {\it is} important in the
calculation of the lifetime of the neutron.

Another important result is the total rate of decay Eq. (27). This
expression determines the total probability that the particle decays, and as
we can see, it depends of the constant of Fermi $G_{F}$, of the parameter of
left-right asymmetry $\lambda $, of $\rho $ that is originated by the strong
interaction, and of $\epsilon _{m}$.

As it was already commented, the parameter $\lambda$ does not alter the
form of the spectrum of energy, but we believe that this does play an
important role in the calculation of the lifetime of the neutron, since in
the Eq. (27) the factor $(1+\lambda ^{2})$ appears, that whether $\lambda
=\pm 1$ would have a factor of 2, that would be considerable for the numeric
calculation of the lifetime, besides $\rho $ and $\epsilon _{m}$. Now we
present the calculation of the lifetime of the neutron

\begin{equation}
\tau = \frac{1}{\Gamma}=1060.85 \hspace*{1mm}seg.
\end{equation}

Although this result is very far from the most recent experimental value
about the lifetime of the neutron \cite{Data}, this calculation illustrates
the ease of determining the lifetime once the expression for the total rate
of decay is obtained. Furthermore, we stress the importance implied in
determining more precise analytical expressions on the total rate of decay
and as a consequence of the lifetime, which plays a central role in the
theory of the weak interaction, since it provides a mensuration of the rate
of the constants of coupling $g_{V}/g_{A}$. On the other hand, to have a
reliable value of the lifetime of the neutron is also of great importance in
astrophysics in relation to the problem of the abundance of helium in the
universe.

The numeric result of the correction term $\epsilon_m$ of the Eq. (27),
which is originated when considering the recoil of the proton and the
difference of masses between the proton and the neutron is:

\begin{equation}
\epsilon_m=\frac{1}{2}(1-z^2)=0.0013755\pm 5.9477\times 10^{-7}.
\end{equation}

\noindent Comparing our result with that corresponding to the radiative
corrections \cite{Augusto1}, we find that both results are of the same
order of magnitude and they coincide up to the fourth decimal digit, which
implies that our result is good, since it compares an approach to first
order with recoil of the proton and difference of masses $(m_{p}\neq m_{n})$,
with a radiative correction. The other correction terms in the Eq. (27)
have rejected their contribution since it is very small.

Modifying the Eq. (27) by means of the approaches that commonly are made,
that is, to take the limit $V-A$ and to make the masses of the proton and of
the neutron the same $m_{p}=m_{n}$, we obtain:

\begin{equation}
\Gamma=\frac{G^2_Fm^5_n\epsilon^5_m}{60\pi^3}(1+3\rho),
\end{equation}

\noindent which is the result that commonly appears in the literature, which
implies that our calculations are consistent with the theory.

\newpage

\begin{center}
{\large {\bf Appendix}}
\end{center}

\vspace{1cm}

This appendix contains the calculations corresponding to the total decay
rate, for the case when the mass of the electron is different from zero,
that is, $y=\frac{m_{e}}{m_{n}}\neq 0$ which corresponds to a more complete
analysis. We start from the Eq. (22) but now conserving all the terms with $y$.

After a simple algebra, it is found that the new expression for the spectrum
of energy of the electrons is:

\begin{equation}
\frac{d\Gamma}{d\epsilon}=\frac{2G^2_F m^5_n}{(2\pi)^3}\frac{\eta}{b^3_0(c_0+\epsilon)^3}\sum^5_{n=0}F^{\prime}_n\epsilon^n,
\end{equation}

\noindent where:

\begin{eqnarray}
F^{\prime}_0&=&P^{\prime}\epsilon^{^{\prime}2}_m,  \nonumber \\
F^{\prime}_1&=&Q^{\prime}\epsilon^{^{\prime}2}_m-2P^{\prime}\epsilon^{\prime}_m,  \nonumber \\
F^{\prime}_2&=&R^{\prime}\epsilon^{^{\prime}2}_m-2Q^{\prime}\epsilon^{\prime}_m+P^{\prime},  \nonumber \\
F^{\prime}_3&=&S^{\prime}\epsilon^{^{\prime}2}_m-2R^{\prime}\epsilon^{\prime}_m+Q^{\prime},  \nonumber \\
F^{\prime}_4&=&-2S^{\prime}\epsilon^{\prime}_m+R^{\prime}  \nonumber \\
F^{\prime}_5&=&S^{\prime},  \nonumber
\end{eqnarray}

\begin{eqnarray}
P^{\prime}&=&(-A_\rho A_\lambda -B_\rho B_\lambda +A_\lambda C_\rho z)y^2
+(-A_\rho A_\lambda - B_\rho B_\lambda +A_\lambda C_\rho z)y^4  \nonumber \\
&& +\frac{4}{3}(-A_\rho A_\lambda +B_\rho B_\lambda)\epsilon^{\prime}_m y^2 \nonumber \\
Q^{\prime}&=&2A_\rho A_\lambda -A_\lambda C_\rho z +(-A_\rho A_\lambda -
B_\rho B_\lambda)\epsilon^{\prime}_m +(\frac{14}{3}A_\rho A_\lambda +\frac{2}{3}B_\rho B_\lambda -3A_\lambda C_\rho z)y^2  \nonumber \\
&& +2A_\rho A_\lambda y^4+(-A_\rho A_\lambda - B_\rho
B_\lambda)\epsilon^{\prime}_m y^2,  \nonumber \\
R^{\prime}&=&-6A_\rho A_\lambda+2B_\rho B_\lambda+2A_\lambda C_\rho z+\frac{2}{3}(A_\rho A_\lambda+B_\rho B_\lambda)\epsilon^{\prime}_m +(-6A_\rho
A_\lambda+2B_\rho B_\lambda)y^2,  \nonumber \\
S^{\prime}&=&\frac{16}{3}A_\rho A_\lambda-\frac{8}{3}B_\rho B_\lambda. \nonumber
\end{eqnarray}

\noindent As it can be observed, it has increased the number of terms in
(31), which makes a more laborious algebra.

\noindent What follows is to determine the total rate of decay; for this we
integrate the Eq. (31) with regard to the variable $\epsilon $:

\begin{equation}
\Gamma=\frac{2G^2_F m^5_n}{(2\pi)^3b^3_0}\sum^5_{n=0}F^{\prime}_nI^{\prime}_n,
\end{equation}

\noindent where

\[
I_n=\int^{\epsilon^{\prime}_m}_{\epsilon_0}\eta\frac{\epsilon^n}{(c_0+\epsilon)^3}d\epsilon, \hspace*{3mm} n=0,1,2,... ,5, \hspace{3mm}
\epsilon_0=m_e/m_n=y, \hspace{3mm} \epsilon^{\prime}_m=\epsilon_1=\frac{1}{2}
(1+y^2-z^2). 
\]

\noindent To evaluate this type of integral, we carry out the following
variable change:

\noindent we define \hspace*{1cm} 
\[
\alpha =c_{0}+\epsilon ,\hspace*{5mm}d\alpha =d\epsilon \hspace*{5mm}\mbox{y}
\hspace*{5mm}\alpha _{0}=-\frac{1}{2}(1-y^{2}),\hspace*{5mm}\alpha _{1}=-
\frac{1}{2}z^{2}, 
\]

\noindent so that we obtain

\[
I^{\prime}_n=\int^{\alpha_1}_{\alpha_0}\frac{1}{\alpha^3}(\alpha-c_0)^n 
\sqrt{R(\alpha)} d\alpha, 
\]

\noindent with \hspace*{1cm} 
\[
R=C\alpha^2 +B\alpha +A, \hspace*{5mm} C=1, \hspace*{5mm} B=-2c_0, \hspace*{5mm} A=c^2_0 -y^2. 
\]

\noindent Following a similar procedure as in the subsection C

\begin{equation}
\sum^5_{n=1}F^{\prime}_nI^{\prime}_n=\sum^2_{n=-3}G^{\prime}_nJ^{\prime}_n,
\end{equation}

\noindent where the new integrals $J_{n}^{\prime }$ are:

\begin{eqnarray}
J^{\prime}_{-3}&=&\int^{\alpha_1}_{\alpha_0}\frac{1}{\alpha^3}\sqrt{R}
d\alpha = [\frac{2p}{q^2(1+y^2-2\epsilon^{\prime}_m)} -\frac{2}{(1+y^2-2\epsilon^{\prime}_m)^2}] \sqrt{R}+\frac{4y^2}{q^3}L_2,  \nonumber \\
J^{\prime}_{-2}&=&\int^{\alpha_1}_{\alpha_0}\frac{1}{\alpha^2}\sqrt{R}
d\alpha=\frac{2}{(1+y^2 - 2\epsilon^{\prime}_m)} \sqrt{R}-\frac{p}{q}L_2+L_1,
\nonumber \\
J^{\prime}_{-1}&=&\int^{\alpha_1}_{\alpha_0}\frac{1}{\alpha}\sqrt{R} d\alpha=
\sqrt{R}-\frac{1}{2}qL_2+\frac{1}{2}pL_1,  \nonumber \\
J^{\prime}_{0}&=&\int^{\alpha_1}_{\alpha_0}\sqrt{R}d\alpha=\frac{1}{2}
\epsilon^{\prime}_m\sqrt{R}-\frac{1}{2}y^2L_1,  \nonumber \\
J_1&=&\int^{\alpha_1}_{\alpha_0}\alpha \sqrt{R}d\alpha=[-\frac{1}{3}y^2 - 
\frac{1}{4}\epsilon^{\prime}_m+\frac{1}{3}\epsilon^{^{\prime}2}_m-\frac{1}{4}
y^2\epsilon^{\prime}_m]\sqrt{R}+\frac{1}{4}py^2L_1,  \nonumber \\
J^{\prime}_2&=&\int^{\alpha_1}_{\alpha_0}\alpha^2 \sqrt{R}d\alpha=[\frac{1}{3}y^2
+\frac{1}{3}y^4 +\frac{1}{8}\epsilon^{\prime}_m - \frac{1}{3}
\epsilon^{^{\prime}2}_m +\frac{1}{4}\epsilon^{^{\prime}3}_m +\frac{1}{8}y^2
\epsilon^{\prime}_m+\frac{1}{8}y^4 \epsilon^{\prime}_m-\frac{1}{3}y^2
\epsilon^{^{\prime}2}_m]\sqrt{R}  \nonumber \\
&& -\frac{1}{32}(5p^2-q^2)y^2L_1.  \nonumber
\end{eqnarray}

\noindent To solve this type of integral is more laborious than those
obtained in the subsection C. In these new integrals we define the following

\[
p=1+y^2, \hspace*{3mm} q=1-y^2, \hspace*{3mm} \epsilon^{\prime}_m=\epsilon_m
+\frac{1}{2}y^2, \hspace*{3mm} z^2=1+y^2-2\epsilon^{\prime}_m, \hspace*{5mm}
L_1=log\frac{p-z^2+2\sqrt{R}} {2y}, \hspace*{5mm} 
\]
\[
L_2=log \frac{q^2-z^2p+2q\sqrt{R}}{2yz^2},\hspace*{5mm} \sqrt{R}=\frac{1}{2}
\sqrt{z^4-2z^2p +q^2}=\sqrt{\epsilon^{^{\prime}2}_m-y^2}. 
\]

\noindent The corresponding coefficients $G_{n}^{\prime }$ for this case are:

\begin{eqnarray}
G^{\prime}_{-3}&=&\frac{1}{24}(1+y^2-2\epsilon^{\prime}_m)^3(1-y^2)^2(A_\rho
A_\lambda +B_\rho B_\lambda),  \nonumber \\
G^{\prime}_{-2}&=&(1+y^2-2\epsilon^{\prime}_m)[(A_\rho A_\lambda +B_\rho
B_\lambda)\{(1-y^2)(\frac{1}{6}+\frac{1}{6}y^2 +\frac{1}{6}y^4-\frac{5}{12}
\epsilon^{\prime}_m  \nonumber \\
&& +\frac{1}{6}\epsilon^{^{\prime}2}_m-\frac{1}{4}y^2\epsilon^{\prime}_m-
\frac{2}{3}y^4\epsilon^{\prime}_m) +\frac{1}{3}y^6-\frac{2}{3}
y^6\epsilon^{\prime}_m+\frac{1}{3}y^2\epsilon^{^{\prime}2}_m \} +\frac{1}{4}
A_\lambda C_\rho z(1-y^2)(1+y^2-2\epsilon^{\prime}_m)],  \nonumber \\
G^{\prime}_{-1}&=&A_\rho A_\lambda(\frac{2}{3}+2y^2+2y^4+\frac{2}{3}y^6-
\frac{5}{2}\epsilon^{\prime}_m +2\epsilon^{^{\prime}2}_m+\frac{2}{3}
\epsilon^{^{\prime}3}_m-5y^2\epsilon^{\prime}_m-\frac{5}{2}
y^4\epsilon^{\prime}_m+2y^2\epsilon^{^{\prime}2}_m)  \nonumber \\
&& +B_\rho B_\lambda(-\frac{1}{3}-y^2-y^4-\frac{1}{3}y^6+\frac{3}{2}
\epsilon^{\prime}_m -2\epsilon^{^{\prime}2}_m+\frac{2}{3}\epsilon^{^{\prime}3}_m
+3y^2\epsilon^{\prime}_m+\frac{3}{2}y^4\epsilon^{\prime}_m-2y^2
\epsilon^{^{\prime}2}_m)  \nonumber \\
&& +A_\lambda C_\rho z(\frac{2}{3}+y^2-\frac{1}{2}y^4-4\epsilon^{\prime}_m+2
\epsilon^{^{\prime}2}_m), \\
G^{\prime}_0&=&A_\rho A_\lambda(\frac{10}{3}+\frac{22}{3}y^2+\frac{10}{3}
y^4-9\epsilon^{\prime}_m +4\epsilon^{^{\prime}2}_m-9y^2\epsilon^{\prime}_m) 
\nonumber \\
&& +B_\rho B_\lambda(-\frac{8}{3}-\frac{14}{3}y^2-\frac{8}{3}
y^4+7\epsilon^{\prime}_m -4\epsilon^{^{\prime}2}_m+7y^2\epsilon^{\prime}_m)
+ A_\lambda C_\rho z(3+y^2-4\epsilon^{\prime}_m),  \nonumber \\
G^{\prime}_1&=&A_\rho A_\lambda(\frac{22}{3}+\frac{22}{3}y^2-10\epsilon^{\prime}_m)
+B_\rho B_\lambda(-\frac{14}{3}-\frac{14}{3}y^2+6\epsilon^{\prime}_m) +2A_\lambda C_\rho z,  \nonumber \\
G^{\prime}_2&=&\frac{16}{3}A_\rho A_\lambda-\frac{8}{3}B_\rho B_\lambda. 
\nonumber
\end{eqnarray}

\noindent The following step is to carry out the product of the integrals
with their respective coefficients to determine explicitly (33), and after
carrying out the pertinent adjustments and substituting this result in (32),
the new expression for the total decay rate is:

\begin{eqnarray}
\Gamma&=&\frac{G^2_Fm^5_n}{4\pi^3}(1+\lambda^2)\{(1+\rho^2)[\{1-3\epsilon^{\prime}_m
+\frac{4}{3}\epsilon ^ {^{\prime}2}_m +\frac{2}{3}
\epsilon^{^{\prime}3}_m+y^2(\frac{1}{6}-\frac{1}{3}\epsilon^{\prime}_m+
\frac{7}{3}\epsilon^{^{\prime}2}_m)  \nonumber \\
&& +y^4(\frac{7}{6}-\frac{7}{6}\epsilon^{\prime}_m)-\frac{1}{12}y^6 \}\sqrt{R}
+\{-\frac{1}{2}+2\epsilon^{\prime}_m-2\epsilon^{^{\prime}2}_m+y^4(\frac{1
}{2}-2\epsilon^{\prime}_m+2\epsilon^{^{\prime}2}_m)\} L_2  \nonumber \\
&& +\{\frac{1}{2}-2\epsilon^{\prime}_m+2\epsilon^{^{\prime}2}_m+y^4(-\frac{1}{2}
-2\epsilon^{\prime}_m+2\epsilon^{^{\prime}2}_m)\} L_1] \\
&& +z(1-\rho^2)[\{2-4\epsilon^{\prime}_m +\frac{2}{3}\epsilon^{^{\prime}2}_m
+y^2(-\frac{5}{6}+\frac{5}{3}\epsilon^{\prime}_m)-\frac{1}{3}y^4\} \sqrt{R} \nonumber \\
&&+\{-1+3\epsilon^{\prime}_m-2\epsilon^{^{\prime}2}_m+y^2\epsilon^{\prime}_m(1-2\epsilon^{\prime}_m)
+y^4(-\frac{1}{2}+\epsilon^{\prime}_m)\} L_2 \nonumber \\
&&+\{1-3\epsilon^{\prime}_m+2\epsilon^{^{\prime}2}_m+y^2(-1+2\epsilon^{\prime}_m)+y^4(\frac{1}{2}
+\epsilon^{\prime}_m)\} L_1]\}.  \nonumber
\end{eqnarray}

Comparing this last equation for the total decay rate with that
corresponding to the case when the mass of the electron is zero, that is, $y=\frac{m_{e}}{m_{n}}=0$,
we can see the simple and compact form that it leads
to taking $y=0$ in the Eq. (27), however, the Eq. (35), although it does not
present a very aesthetic form, is very complete.

An important observation of (35) is that when we consider $y=0$, we obtain
exactly the Eq. (27), which means that both results are consistent with the
theory.

It is necessary to mention that to be able to analyze how much the powers of 
$y$ will contribute, it is necessary to rewrite (35) in a simpler form, for
this we must express $\sqrt{R}$, $L_{1}$, and $L_{2}$ as a development in
series of powers. But we should take care when carrying out these
developments since they are very sensitive to any change, which is why they
should be analyzed carefully.

\hspace{2cm}

\begin{center}
{\bf Acknowledgments}
\end{center}

This work was supported in part by {\it Consejo Nacional de Ciencia y
Tecnolog\'{i}a} (CONACyT), {\it Sistema Nacional de Investigadores} (SNI)
(M\'{e}xico) and Programa de Mejoramiento al Profesorado (PROMEP). A.G.R.
would like to thank the organizers of the Summer School in Particle Physics
and Sixth School on non-Accelerator Astroparticle Physics 2001, Trieste
Italy for their hospitality. The authors would also like to thank Anna Maria
D'Amore for revising the manuscript.

\newpage

\begin{center}
{\bf FIGURES}
\end{center}

\vspace{5mm}

\noindent {\bf Fig. 1} Diagram of Feynman for the decay of the neutron
$n\rightarrow p+e^- +\bar\nu_e$.

\bigskip

\noindent {\bf Fig. 2} Spectrum of energy of the electrons for $\lambda=1$
(right currents).

\bigskip

\noindent {\bf Fig. 3} Spectrum of energy of the electrons for $\lambda=-1$
(left currents).

\bigskip

\noindent {\bf Fig. 4} Spectrum of energy of the electrons for four values
different from the parameter of asymmetry $\lambda =1,-1,0.9,-0.9$.

\end{document}